\documentclass{amsart}

\theoremstyle{definition}

\theoremstyle{remark}

\usepackage{graphicx}

\usepackage{color}
\usepackage{subfigure}
\usepackage{ulem}

\usepackage{amsmath,epsf}

\numberwithin{equation}{section}

\usepackage[dvipsnames,svgnames,table]{xcolor}

\newcommand{\ii}{\mathrm{i}}
\newcommand{\dd}{\mathrm{d}}
\newcommand{\ee}{\mathrm{e}}
\newcommand{\bv}[1]{\mathrm{\mathbf{#1}}}

\newcommand{\eps}{\varepsilon}
\newcommand{\grad}{\nabla}

\newcommand{\ft}[1]{\widehat{#1}}
\newcommand{\ol}[1]{\overline{#1}}

\newcommand{\pdv}[2]{\frac{\partial #1}{\partial #2}}
\newcommand{\ddv}[2]{\frac{\dd #1}{\dd #2}}

\NewDocumentCommand{\SubAlign}{m o}{
        \begin{subequations} 
            \begin{align} 
                #1
            \end{align}    
            \IfValueTF{#2}{\label{#2}}{} 
    \end{subequations}
    }

\title{Spiral wave solutions in water waves}

\author{Mark J. Ablowitz}
\address{Department of Applied Mathematics, University of Colorado, Boulder, Colorado 80309-0526}
\email{mark.ablowitz@colorado.edu}
\thanks{}

\author{Justin T. Cole}
\address{Department of Mathematics, University of Colorado Colorado Springs, 80918}
\email{jcole13@uccs.edu}

\author{Sean D. Nixon}
\address{Department of Applied Mathematics, University of Colorado, Boulder, Colorado 80309-0526}
\email{Sean.Nixon@colorado.edu}

\begin{document}
\maketitle

\begin{abstract}
Spiral wave solutions are found in linear and weakly nonlinear irrotational water wave equations. 
These unsteady spiral waves evolve from suitable initial conditions; they are not induced by external forcing. 
In the linear case, a long-time asymptotic result is obtained via the method of stationary phase. 
The asymptotic approximation is found to be in good agreement with the exact solution and reveals hyperbolic spiral structure. Numerical simulations 
show that these spiral waves persist in the presence of weak nonlinearity. While spiral solutions are frequently found in excitable media governed by reaction-diffusion systems, they 
comprise a new class of interesting two space one time dimensional solutions in fundamental linear and nonlinear dispersive wave systems.
\end{abstract}

\section{Introduction}
The analytical study of classical water waves is one of the oldest subjects in applied mathematics; 
its origins date back to founders of calculus and differential equations: 
Newton, Bernoulli, Euler, Laplace, Lagrange, Cauchy, Airy, Stokes amongst many others 
Finding and understanding the properties of solutions to the water wave equations have been central in the study of wave phenomena. 
In this article we study irrotational water waves with localized initial data on an unbounded domain. 
In the case of linear waves, the problem can be solved via Fourier transforms where we need 
to obtain the underlying dispersion relation (or wave frequency) associated with a typical wave.

Nonlinear waves are more difficult; in this case various asymptotic and certain linearization techniques play an important role.
In weakly nonlinear deep water waves, Stokes \cite{Stokes1847}  found a relationship between the frequency and amplitude of the dominant Fourier mode of periodic traveling waves. 
Over one hundred years later Benjamin and Feir \cite{BF1967} found that these waves are unstable. 
Soon afterwards, by allowing the envelope of the wave to vary slowly in space and time Zakharov \cite{Zakharov1968} and Benney, 
Roskes \cite{BR1969} showed that the complex amplitude of the envelope satisfies two-space, one-time dimensional nonlinear Schrödinger (NLS)-type equations; 
the equation Benney-Roskes found  is transformable to what is often called a Davey-Stewartson equation \cite{DS1974},\cite{AblCl1991}. 
In shallow water, there are a number of interesting equations and corresponding solutions; the equations include the Boussinesq \cite{boussinesq1871} 
and Benney-Luke \cite{BL1964} equations. 
From these equations asymptotic reductions lead to the unidirectional Korteweg–de Vries (KdV) \cite{KdV1895} equation in one-space, 
one-time dimension and the Kadomtsev–Petviashvili (KP) equation \cite{KP1970}, \cite{AblCl1991} in two-space, one-time dimension. 

Remarkably, some of these equations in unbounded domains with rapidly decaying data have mulit-soliton solutions, 
an infinite number of conserved quantities  and  can be linearized: for example the KdV equation \cite{GGKM1967}, the KP equation (see e.g. \cite{ AblCl1991}), and the one-space, one-time dimensional NLS equation \cite{ ZS1972}. 
It is also noteworthy that resonant three-wave and six-wave interaction equations are also in this class of integrable systems \cite{ ZakMan1975}, \cite{Kaup1976},\cite{AblHab1975}. 
Such equations arise from classical water waves; see e.g. \cite{AblLuoMuss2023} and references therein.

As indicated above, the study of special solutions of water waves has been a major topic over the years; see e.g. \cite{HHSTWWW2022} for a review of the mathematical theory of steady water waves. 
Some of the areas that have attracted substantial interest are: Stokes waves near maximum height \cite{Stokes1880},\cite{LHMF1977},\cite{AFT1982},\cite{DyLushKorot2013};
investigations of solitary and periodic waves including existence, exact solutions and computation; \cite{Beale1977}, \cite{Strauss1977}, \cite{AT1981}, \cite{Groves2004} ,\cite{BergerMilew2000}.

In this article, we show that there is a novel class of two space one time dimensional unsteady spiral wave solutions to the linear and weakly nonlinear irrotational water wave equations. 
These equations are part of a class of purely dispersive wave systems which admit spiral wave solutions. 
Spiral waves are frequently encountered in excitable media such as occurs in reaction-diffusion systems; see e.g. \cite{NicolisPrigogine1977}, 
electric transport systems \cite{BodePurwins1995} and disease spread \cite{CapassoKunisch1988}.
They can also be found in a class of galaxies \cite{LinShu1964}, \cite{Shu2016}, 
optically active crystals \cite{SchellBloembergen1978} and more recently they were found in tunable circular Pearcey beams 
\cite{Chen_etal2021} and wavepacket rotation in symmetry-broken photonic lattices \cite{Liu2021}. 

Spiral waves are not commonly found in fundamental dispersive systems. 
Motivated by our recent studies of Klein-Gordon equations which are related to massive Dirac systems \cite{AblCoNix2025}, 
we have found unsteady spiral waves in linear and weakly nonlinear irrotational water waves.
To our knowledge there are no earlier analytical  studies of the classical water wave equations that feature 
such spiral wave solutions. Perhaps this is due to the fact that these spirals are  a two-space one time dimensional phenomena which evolve from a certain class of initial conditions (ICs); see equations \eqref{Eq: SpiralIC}.
While such ICs are elementary, they are not obvious; we were led to these ICs by studying topological wave dynamics in linear and nonlinear optics \cite{AblCoNix2025}. 
Finally, it should be noted that numerous photographs were taken of spiral-type waves in the various oceanic regions by astronauts in early space flight missions. 
These photographs were carefully studied by Munk et al. \cite{Munk2000}; they attributed the spiral waves to horizontal shear instability modified by rotational effects. 
Nevertheless, the fact that the water wave equations admit spirals without external forcing is helpful in maintaining such structures.

This paper is organized as follows. In section 2 the governing equations are given; the linear and weakly nonlinear systems are derived. 
In section 3 the spiral solutions are shown to evolve from a class of initial data. The linear water wave equations are solved via Fourier transforms and an asymptotic approximation is obtained using stationary phase methods. 
The approximation is compared  with the exact solution with good agreement. 
In section 4 the weakly nonlinear system is  numerically studied. 
The spiral waves are found to persist for moderate size of nonlinearity. Concluding remarks are made in section 5.

\section{Governing Equations}
\label{Sec: Model}
The free surface irrotational water wave equations with a flat bottom, depth $h$, 
 satisfy the following equations in the absence of surface tension
\begin{itemize} 
  \item [] 
  Ideal flow
    \begin{equation}
        \label{eq:laplace}
      \nabla^2 \phi = 0, \qquad  -h<z<\eta({\bf r},t),
    \end{equation}
    
  \item []
  No flow through bottom
    \begin{equation}
        \label{eq:bottom}
      \frac{\partial\phi}{\partial z} = 0, \qquad \text{~on~}  z=-h,
    \end{equation}
    
  \item []  
  Bernoulli or pressure equation
    \begin{equation}
        \label{eq:bernoulli}
      \frac{\partial\phi}{\partial t} +
        \frac{1}{2}|{\nabla}\phi |^2 + \frac{1}{2}\left(  \frac{\partial \phi}{\partial z}  \right)^2 + 
        g\eta = 0 \qquad \text{~on~} z=\eta({\bf r},t),
    \end{equation}
  \item []
  Kinematic boundary condition
    \begin{equation}
        \label{eq:kinematic}
      \frac{\partial\phi}{\partial z} = \frac{\partial\eta}{\partial
        t} + \nabla \phi \cdot \nabla \eta, \qquad \text{~on~}  z=\eta({\bf r},t) \;,
    \end{equation}
\end{itemize}
where $\nabla = \partial_x^2 + \partial_y^2 $ is the horizontal gradient operator and $g$ is the acceleration due to gravity. 
These four equations constitute the classical equations for irrotational water waves. 
Here, the unknowns are: $\phi({\bv{r}},z,t)$ the velocity potential; $\eta({\bf r},t)$ the surface wave elevation, 
${\bf r}=(x,y)$ is the horizontal coordinate, $z$ the vertical coordinate and $t$ is time. 
This is a free-boundary problem for the unknowns $\phi({\bf r},z,t)$ and $\eta({\bf r},t)$.
 
In \cite{AFM2006}, the water wave problem was reformulated as a nonlocal differential-integral system for two surface unknowns,
$\eta({\bf r},t)$ and $q=q({\bf r},t)= \phi({\bf r},\eta({\bf r},t),t)$. 
In the plane, the equations are given by 
\begin{subequations}
  \begin{equation}
    \iint_{\mathbb{R}^2} d {\bf r} e^{ - i {\bf k} \cdot {\bf r} }
      \left( i \eta_t \cosh[k (\eta+h)] 
      -  \frac{{\bf k} \cdot \nabla q}{k}  \sinh \left[ k (\eta+h)\right]\right) = 0 \;,
        \label{Eq: EtaQEquation1}
  \end{equation}
  \begin{equation}
    q_t + \frac{1}{2} | \nabla q|^2 + g\eta - \frac{(\eta_t + \nabla q \cdot \nabla \eta)^2} {2(1+|\nabla \eta|^2)}=0\;,
      \label{Eq: EtaQEquaton2}
  \end{equation}
    \label{Eqs: EtaQEquatons}
\end{subequations}
where $k = | {\bf k}| = \sqrt{k_x^2 + k_y^2} \ge 0$.
It is assumed that $\eta, \nabla q, q_t$ decay rapidly to zero at infinity. 
Equation \eqref{Eq: EtaQEquaton2} is Bernoulli's equation on the free surface. 
The benefit of this formulation is that it provides an explicit formulation of the surface variables.
The nonlocal formulation is particularly useful for asymptotic calculations such as the ones in this paper. 
In the infinite depth limit, $h \to \infty$, Eq.~\eqref{Eq: EtaQEquation1} reduces to
\begin{equation}
  \iint_{\mathbb{R}^2} d {\bf r} e^{ - i {\bf k} \cdot {\bf r} } e^{k\eta}
    \left( i \eta_t    -  \frac{ {\bf k}}{k} \cdot \nabla q  \right) = 0 \;.
    \label{Eq: EtaQEquation1a}
\end{equation}
We consider the weakly nonlinear waves case for which it is convenient to let  $\eta \to \eps \eta, q \to \eps q$ and assume $|\eps| \ll1$. 
Doing so and expanding the hyperbolic functions in Eq.~\eqref{Eq: EtaQEquation1} to order $\epsilon$ we find: (i) for finite depth,
\begin{equation}
  \iint_{\mathbb{R}^2} d {\bf r} e^{-i {\bf k} \cdot {\bf r} } \left( i \eta_t (1 + \eps k \tanh (k h) \eta  ) 
    -  \frac{ {\bf k}}{k} \cdot \nabla q  (\tanh (k h) +  \eps  k \eta  )  + \ldots \right) =0 \;,
    \label{Eq: NLEtaIntegral}
\end{equation}
and (ii) for infinite depth, or $h \to \infty$,
\begin{equation}
  \iint_{\mathbb{R}^2} d {\bf r} e^{-i {\bf k} \cdot {\bf r} } \left( (i \eta_t -\frac{ {\bf k}}{k} \cdot \nabla q)  (1+  \eps k \eta)+ \cdots \right) =0 \;.
    \label{Eq: EtaQEquation1A}
\end{equation}
Note that these two equations differ only by simple factors.
The free surface Bernoulli equation \eqref{Eq: EtaQEquaton2} is unchanged regardless of finite or infinite depth; 
this equation to order $\eps$ reads
\begin{equation}
  q_t=-g\eta  +\frac{\eps}{2}(\eta_t^2-|\nabla q|^2)+ \ldots\;. 
    \label{Eq: EtaQEquaton2A}
\end{equation}
The weakly nonlinear equations are considered  in more detail in Section \ref{Sec: weakNL}.

We define the Fourier and inverse Fourier transforms as 
\begin{subequations}
  \begin{equation}
    \hat{f}({\bf k})= \frac{1}{2\pi } \iint_{\mathbb{R}^2} d {\bf r}  {f}({\bf r} )e^{-i {\bf k} \cdot {\bf r} } , \\
      \label{FT}
  \end{equation}
  \begin{equation}
    f({\bf r})= \frac{1}{2\pi } \iint_{\mathbb{R}^2} d {\bf k}  \hat{f}({\bf k} )e^{i {\bf k} \cdot {\bf r} } .
      \label{IFT}
  \end{equation}
\end{subequations}
Then from equations \eqref{Eqs: EtaQEquatons} the linearized equations satisfy, respectively,
\begin{equation}
  i\hat{\eta}_t -\frac{ {\bf k}}{k} \cdot \widehat{ \nabla q} \tanh (k h)=0  ,
    \label{Eq: LinearWW1}
\end{equation}
\begin{equation}
  \widehat{\nabla q}_t =-g\widehat{ \nabla \eta} ,
    \label{Eq: LinearWW2}
\end{equation}
where $\widehat{ \nabla f} =  i {\bf k} \hat{f}$. Differentiating Eq.~\eqref{Eq: LinearWW1} and combining these equations leads to 
\begin{equation}
  \hat{\eta}_{tt}+ \omega^2(k) \hat{\eta}=0,
    \label{Eq: LinearGeneralOmega}
\end{equation}
where $\omega(k) $ is the linear dispersion relation for two-dimensional water waves in finite depth given by
\begin{equation}
 \omega^2(k)=  gk\tanh (k h), ~~~~ k^2= k_x^2+k_y^2.
 \label{Eq: OmegaFiniteDepth}
\end{equation}
In the infinite depth limit, $h \to \infty$, the dispersion relation is given by 
\begin{equation}
  \omega^2(k)=  gk\;. 
    \label{Eq: OmegaDeepWater}
\end{equation}
The surface wave $\eta(x,y,t)$ can be obtained by solving Eq.~\eqref{Eq: LinearGeneralOmega} 
and then taking its inverse Fourier transform. When we describe the {\it exact} solution below, this is what we are referring to.


\section{Linear Spiral Waves}
\label{linear_spiral_sec}

In this section, linear spiral wave solutions are observed for certain initial conditions and analyzed via the method of stationary phase.
In order to investigate spiral solutions, we will solve for the wave elevation $\eta({\bf r},t)$ from equation \eqref{Eq: LinearGeneralOmega} with the following initial conditions 
\begin{equation}
  \eta(x,y,t=0)= xe^{-r^2}, ~~ \eta_t(x,y,t=0)= ye^{-r^2}, ~r^2=x^2+y^2.
    \label{Eq: SpiralIC}
\end{equation}
Physically, this corresponds to spatial and velocity profiles which are odd about the origin and 90 degrees out-of-phase with each other. 
This configuration provides the initial ``twist'' sufficient to generate spiral waves. 
A larger class of initial data of the form $\eta(x,y,t=0)= P_1(x,y) f(r), ~\eta_t(x,y,t=0)= P_2(x,y) f(r)$ where $P_j(x,y)$ are suitable polynomials and $f(r)$ is a rapidly decaying radially symmetric function will also lead to spiral solutions. 
However, the initial condition  above has a rather simple form that achieves our goals.
 
\begin{figure}
  \begin{center}
    \includegraphics[width = 0.95\textwidth]{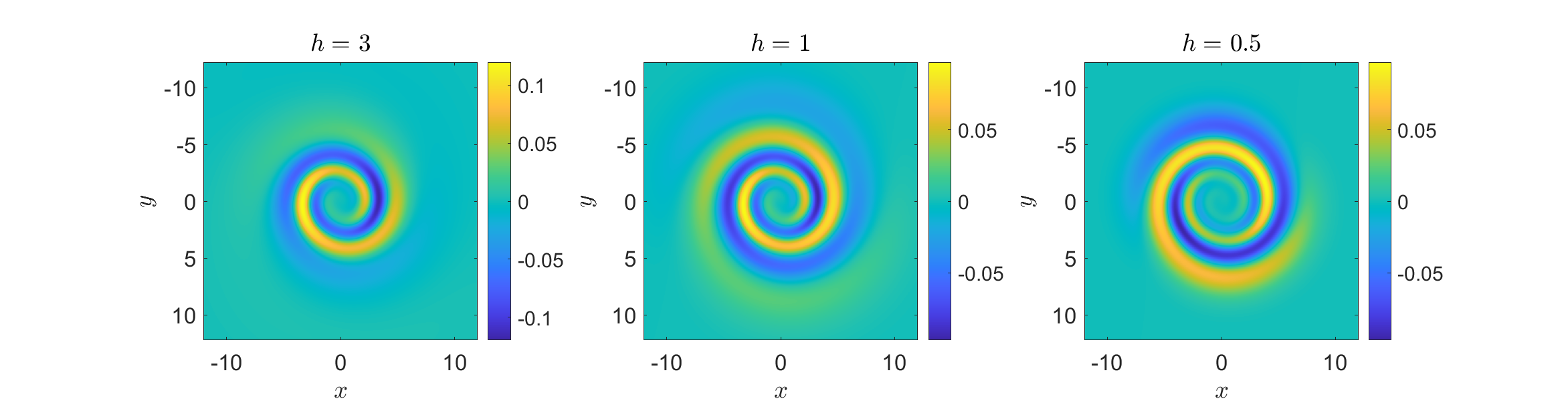}
    \caption{Evolution of the linearized water wave equation \eqref{Eq: LinearGeneralOmega} with dispersion relation \eqref{Eq: OmegaFiniteDepth} for initial condition \eqref{Eq: SpiralIC}.
    Here $t=10$ and $g = 1$.}
        \label{Fig: LinearEvolution}
  \end{center}
\end{figure}

Typical evolutions of the linear water wave equation appear in Figure~\ref{Fig: LinearEvolution} for different fluid depths. 
Examining the differences, the shallower depths have more identifiable spirals, but have lost more amplitude. 
We do not show it here, but the finite depth solutions for $h > 3$ 
are indiscernible from the infinite depth limit dynamics \eqref{Eq: OmegaDeepWater}. 
These results clearly indicate the presence of spiral wave solutions in linear water wave systems. 
We note that these are induced by the initial conditions and not by external forcing.

Having observed these spirals numerically, we seek to describe them analytically. 
Motivated by previous work used to describe spiral motion in a Klein-Gordon equation \cite{AblCoNix2025}, 
we implement stationary phase to describe the structure of these spirals.
The second-order linear water wave equation \eqref{Eq: LinearGeneralOmega} has general solution 
\begin{equation}
    \eta(x,y,t) = I_+(\bv{r},t) + I_-(\bv{r},t)
\end{equation}
where
\begin{align}
    I_{\pm}(\bv{r}, t) &= \frac{1}{2\pi} \iint_{\mathbb{R}^2} \ft{A}_{\pm}(\bv k)\, \ee^{\ii\big(\bv k \cdot \bv r \pm \omega(k) \,t\big)}\dd k_x \, \dd k_y\\
     & =\frac{1}{2\pi} \iint_{\mathbb{R}^2} \ft{A}_{\pm}(\rho,\phi)\, \ee^{\ii\big(\rho \cos \phi\, x + \rho \sin \phi \, y \pm \omega(\rho) \, t \big)} \rho\, \dd \phi\, \dd \rho .\notag
\end{align}
In the polar version of the inverse Fourier transform, we take $k_x = \rho \cos \phi$ and $k_y = \rho \sin \phi$. 
We intentionally take $(\rho, \phi)$ to denote the {\it spectral} polar coordinates to distinguish them from the standard spatial labels $(r, \theta)$. 
Notice that the dispersion relation only depends on the modulus of the Fourier wavenumbers. The coefficients 
$\ft{A}_{\pm}$ are related to the initial conditions by
\begin{equation}
  \ft{A}_{\pm}(\rho  , \phi) = \frac{1}{2}\left( \ft{\eta}(\rho, \phi,0) \pm \frac{1}{\ii \omega(\rho)}\,\pdv{\ft{\eta}}{t}(\rho,\phi,0) \right).
  \label{Fourier_coeff}
\end{equation}
As indicated above, the long time, $t\gg1$, asymptotic approximation of the solutions are 
obtained through a stationary phase analysis. 
This is done by first approximating in $\phi$, then $\rho$. Since the dispersion relation only depends on $\rho$, this approach simplifies the calculation. 
Scaling out time, we define the $O(1)$ variables 
\begin{equation}
  \ol{x} = \frac{x}{t} \quad \ol{y} = \frac{y}{t} \quad \ol{r} = \frac{r}{t}.
\end{equation} 
First, isolate the integral in $\phi$,
\begin{equation}
    J_{\pm}(\rho, \ol{r},\theta, t) = \int_{-\pi}^{\pi} \ft{A}_{\pm}(\rho,\phi)\, \ee^{\ii \big(\rho \cos \phi\, \ol{x} + \rho \sin \phi \, \ol{y} \big)\, t} \dd \phi,
\end{equation}
which has two stationary points of the Fourier phase, $\phi_j$ $j = 1,2$, in the interval $[-\pi ,\pi]$ that satisfy the equation
\begin{equation}
    \tan \phi_{j} = \frac{~\ol{y}~}{\ol{x}},
    \label{stationary_pt_theta}
\end{equation}
where $ \rm{sgn}(\ol{x}) = \rm{sgn}\big(\cos \phi_1\big)$ and $ \rm{sgn}(\ol{y}) = \rm{sgn}\big( \sin \phi_1\big)$ for $\phi_1$, while 
$ \rm{sgn}(\ol{x}) = -\rm{sgn}\big(\cos \phi_2\big)$ and $ \rm{sgn}(\ol{y}) = -\rm{sgn}\big( \sin \phi_2\big)$ for $\phi_2$. 
That is, there are a total of {\it four} stationary points: two for the positive case and two for the negative. 
Also notice that the spectral angle $\phi$ is precisely the spatial polar  angle $\theta$.
The second derivative of the phase is given by
\SubAlign{
   \ddv{^2}{\phi^2}\Big(\rho \cos \phi\, \ol{x} + \rho \sin \phi \, \ol{y} \Big|_{\phi_j}
    & = \rho \big(- \cos \phi_j\, \ol{x} -\sin \phi_j \, \ol{y} \big) \\
    & =(-1 )^{j} \rho\, \ol{r}.
}
Using this, we find the stationary phase approximation 
\begin{equation}
    J_{\pm}(\rho, \ol{r}, \theta, t)  \approx \sqrt{\frac{2 \pi}{\rho  \ol{r} \, t}} \ft{A}_{\pm}(\rho, \phi_1) 
        \ee^{\ii\, \rho \ol{r} \, t - \ii \frac{\pi}{4}} 
        + 
        \sqrt{\frac{2 \pi}{\rho \ol{r} \, t}} \ft{A}_{\pm}(\rho, \phi_2) 
        \ee^{-\, \big(\ii\,\rho \ol{r} \, t - \ii\frac{\pi}{4} \big)} .
\end{equation}

The integral in $\rho$ can now be expressed as
\begin{align}
  I_{\pm}(\ol{r},\theta, t) & \approx  \frac{1}{2\pi}  \int_0^{\infty} J_{\pm}(\rho, \ol{r}, t) \ee^{\pm \ii \omega(\rho) t} \rho~ \dd \rho
    \label{Eq: RhoIntegralDetailed}\\
  & = \ee^{-\ii \frac{\pi}{4}}\frac{1}{\sqrt{2 \pi \ol{r} \, t}}  \int_0^{\infty} \, \ft{A}_{\pm}(\rho, \phi_1)
    \ee^{\ii\,\big(  \rho \,\ol{r}  \pm \omega(\rho)\big) \, t} \sqrt{\rho}~\dd \rho  \notag \\
  & \quad + \ee^{\ii \frac{\pi}{4}}\frac{1}{\sqrt{2 \pi \ol{r} \, t}}  \int_0^{\infty} \, \ft{A}_{\pm}(\rho, \phi_2)
    \ee^{\ii\,\big(- \, \rho \,\ol{r}  \pm \omega(\rho)\big) \, t} \sqrt{\rho}~ \dd \rho . \notag
\end{align}
Defining the phase term
\begin{equation}
    Q_{j}^{\pm}(\rho, \ol{r}) = (-1)^{j+1} \rho \,\ol{r} \pm  \omega\big(\rho\big), ~~~~ j = 1,2
\end{equation}
the stationary points occur for values of $\ol{r}$ that satisfy 
\begin{equation}
  \ddv{Q_{j}^{\pm}}{\rho}  = (-1)^{j+1} \ol{r}  \pm \pdv{\omega}{\rho} = 0 .
    \label{Eq: RhoStationaryPoint}
\end{equation}
For water waves, $\omega'(\rho)>0$ for $\rho>0$, thus solutions to equation \eqref{Eq: RhoStationaryPoint} only occur for 
$j=1$ and the $-$ case or $j=2$ and the $+$ case. So, from the original set of four stationary points in $\phi$, only two are stationary in $\rho$. 
In either case,  the radial stationary point equation is given by 
\begin{equation}
  \ol{r} = \pdv{\omega}{\rho}
    \label{Eq: Rho0General}
\end{equation}
with the same stationary point, $\rho_0$, for both integrals in \eqref{Eq: RhoIntegralDetailed}. 
Checking the second derivative of the phase term, we have
\begin{equation}
    \mathrm{sgn}\left[\pdv{^2 Q_{j}^{\pm}}{\rho^2}\left(\rho_0 \right)\right]  = \mathrm{sgn}\left[\pm \pdv{^2\omega}{\rho^2}\left(\rho_0 \right)\right] =  \mp\,1,
\end{equation}
since $\omega''(\rho)<0$ for water waves. Thus we have the final approximation 
\begin{align}
  \eta(\ol{r},\theta, t) & = I_-(\ol{r},\theta,t)+I_+(\ol{r},\theta,t) 
    \label{Eq: GeneralApproximation}\\
  & \approx \frac{\sqrt{\rho_0}}{ \sqrt{\ol{r} \,|\omega''(\rho_0)|} ~ t} 
    \left( \, \ft{A}_{-}(\rho_0, \phi_1) \ee^{\ii\,\big(  \rho_0 \,\ol{r}  - \omega(\rho_0)\big) \, t} + c.c.\right) , \notag \\
    &  \approx \frac{\rho_0^{3/2}\ee^{-\frac{\rho_0^2}{4}}}{4 \sqrt{\ol{r} \,|\omega''(\rho_0)|} ~ t} 
        \Bigg(   \frac{ \sin \theta}{\omega(\rho_0)} \cos\Big(  \rho_0 \,\ol{r}t  - \omega(\rho_0)t\Big) - \cos \theta  \sin\Big(  \rho_0 \,\ol{r}t  - \omega(\rho_0)t\Big) \Bigg) \notag
\end{align}
where c.c. is complex conjugate,a
valid for $t \gg 1$.
Also note that this unsteady solution is observed to decay like $t^{-1}$ for $\ol{r} = {O}(1)$.
In general, we cannot write down an explicit formula for $\rho_0$ that satisfies Eq.~\eqref{Eq: Rho0General} and then use a numerical approximation. 

\begin{figure}
  \begin{center}
    \includegraphics[width = 0.8\textwidth]{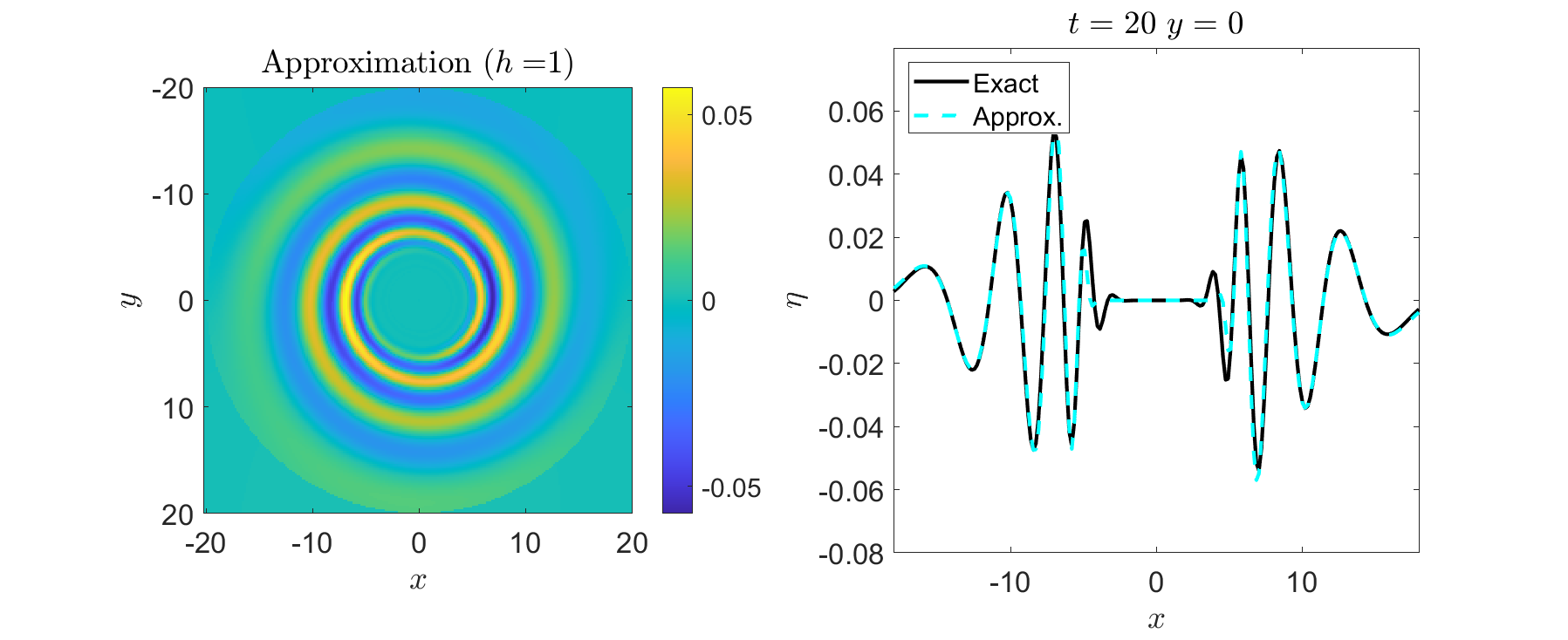}
    \caption{Approximation given by the stationary phase formulae \eqref{Eq: GeneralApproximation} of a finite depth spiral; here $h =1$ and $g=1$. 
    (Left) Contour plot. 
    (Right) Comparison of the exact solution obtained from Eq.~\eqref{Eq: LinearGeneralOmega} 
    and the approximation at cross section $y=0$.}
        \label{Fig: FiniteDepthApproxH1}
  \end{center}
\end{figure}

\begin{figure}
  \begin{center}
    \includegraphics[width = 0.8\textwidth]{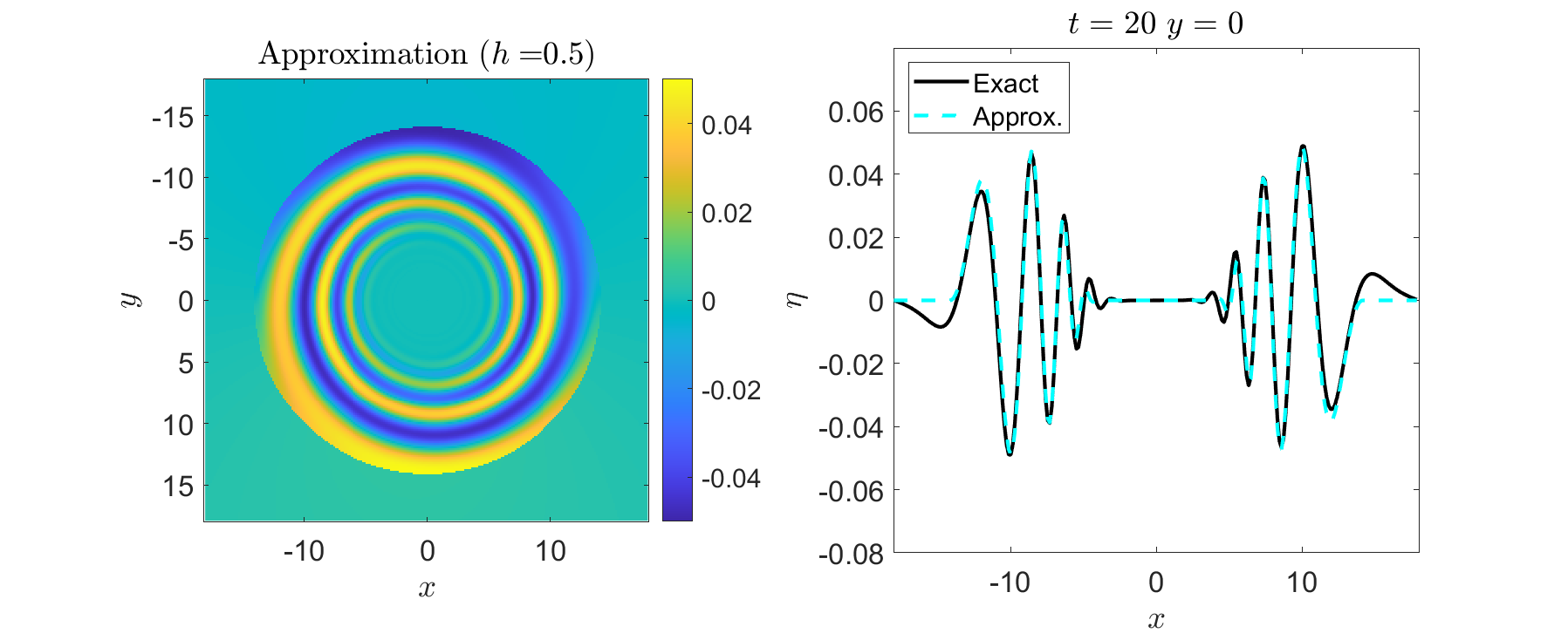}
    \caption{Approximation given by the stationary phase formulae\eqref{Eq: GeneralApproximation} of a finite depth spiral; here $h =0.5$ and $g=1$. (Left) Contour plot. 
    (Right) Comparison of the exact solution obtained from Eq.~\eqref{Eq: LinearGeneralOmega}  
    and the approximation at cross section $y=0$.}
       \label{Fig: FiniteDepthAsymp}
  \end{center}
\end{figure}

A comparison between the exact Fourier solution obtained from Eq.~\eqref{Eq: LinearGeneralOmega} 
and the stationary phase approximation in Eq.~\eqref{Eq: GeneralApproximation} 
is shown in Figures \ref{Fig: FiniteDepthApproxH1} and \ref{Fig: FiniteDepthAsymp} 
for water depths of $h=1$ and $h=0.5$, respectively. 
These solutions were obtained for arbitrary depth and so the modulus of the stationary point, $\rho_0$, is found by numerically solving Eq.~\eqref{Eq: Rho0General}. 
The angular part, for a given $\ol{x},\ol{y}$, is obtained by Eq.~\eqref{stationary_pt_theta}. 
We note that the approximation is limited to the domain $x^2 + y^2 < g h t^2$. Within this region, there is  good agreement with the exact solution. 
Other integral approximations, like steepest descent, are required outside of this region. 
Finally we remark that we find these spirals to be numerically stable under small perturbations.

\subsection{Deep Water}

In the deep water limit, we can get a explicit formula for the spiral shape; here 
$\omega$ is given by equation \eqref{Eq: OmegaDeepWater}, or in polar form 
\begin{equation}
  \omega(\rho) =  g^{1/2} \rho^{1/2}.
\end{equation}
The equation for the stationary points, \eqref{Eq: Rho0General}, now becomes 
\begin{equation}
  \ol{r} = \frac{1}{2} g^{1/2} \rho_0^{-1/2} 
\end{equation}
with solutions
\begin{equation}
    \rho_{0} = \frac{g}{4 \ol{r}^2}.
\end{equation}
Observe  that the spectral modulus point, $\rho_0$, is not equal to spatial modulus $r = \ol{r} t$.

Substituting this into equation \eqref{Eq: GeneralApproximation}, we find 
\begin{equation}
  \eta(\ol{r}, \theta, t) \approx - \frac{g}{t} \,  \frac{1}{ 2 \sqrt{2}~\ol{r}^{3}} \left( \ft{A}_{-}(\rho_0, \phi_1)
        \ee^{-\ii \frac{g}{4 \ol{r} } \, t} + c.c. \right)
\end{equation}
with initial conditions \eqref{Eq: SpiralIC} from Section \ref{Sec: Model}. 
After the Fourier coefficients in Eq.~\eqref{Fourier_coeff} are evaluated we have 
\begin{equation}
   \eta(\ol{r}, \theta,t) \approx \frac{g^2 \ee^{-\frac{g^2}{64 \ol{r}^4}}}{32 \sqrt{2}\, \ol{r}^5 \, t} 
    \left( \frac{2 \ol{r}}{g} \sin(\theta) \cos\left(\frac{g t}{4 \ol{r}} \right) -   \cos(\theta) \sin\left(\frac{g t}{4 \ol{r}} \right)\right) . 
    \label{Eq: InfiniteDepthApprox}
\end{equation}
To obtain this result, recall that the two angular stationary points in Eq.~\eqref{stationary_pt_theta} must have opposite signs. 
This unsteady solution has three distinctive parts: (i) linear decay in time; (ii) 
envelope; and (iii) spiral structure inside the parentheses. 
In regard to the envelope structure, we observe a dominant exponential decay to zero when $\ol{r} \rightarrow 0$, 
but an algebraic (quintic) decay as $\ol{r} \rightarrow \infty$. The spiral structure will be further analyzed in the next section.

\begin{figure}
  \begin{center}
    \includegraphics[width = 0.8\textwidth]{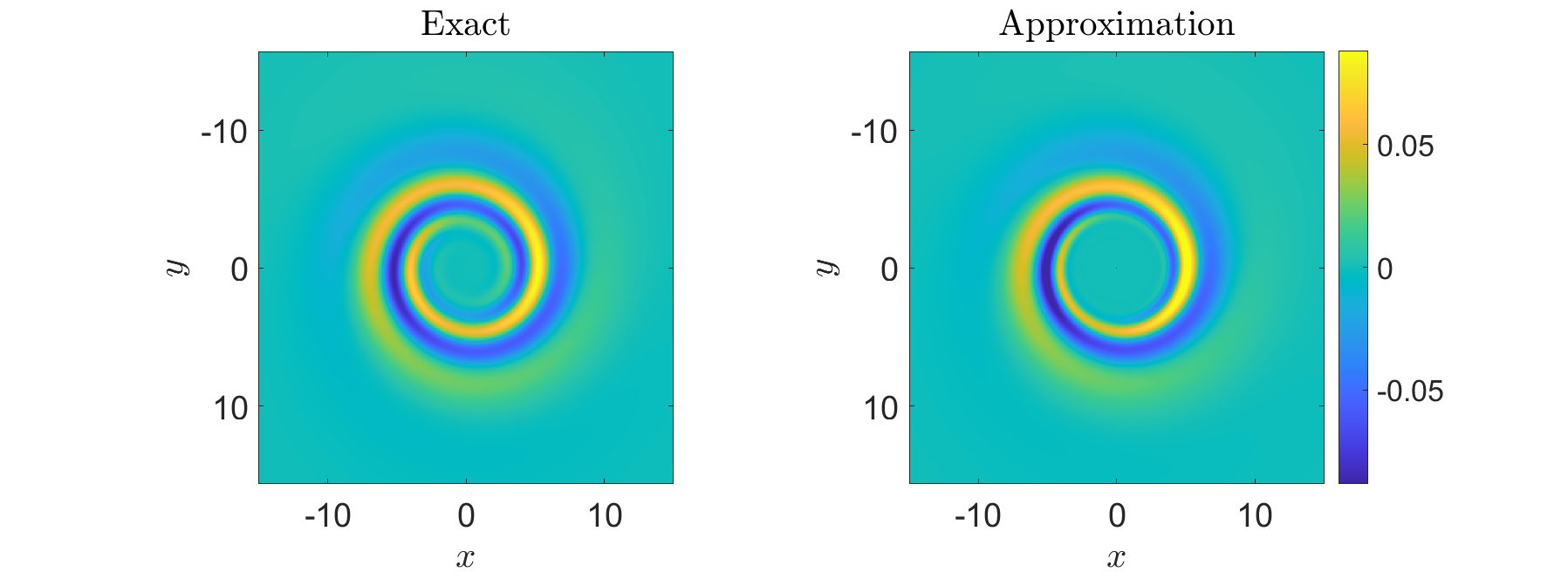}
    \includegraphics[width = 0.8\textwidth]{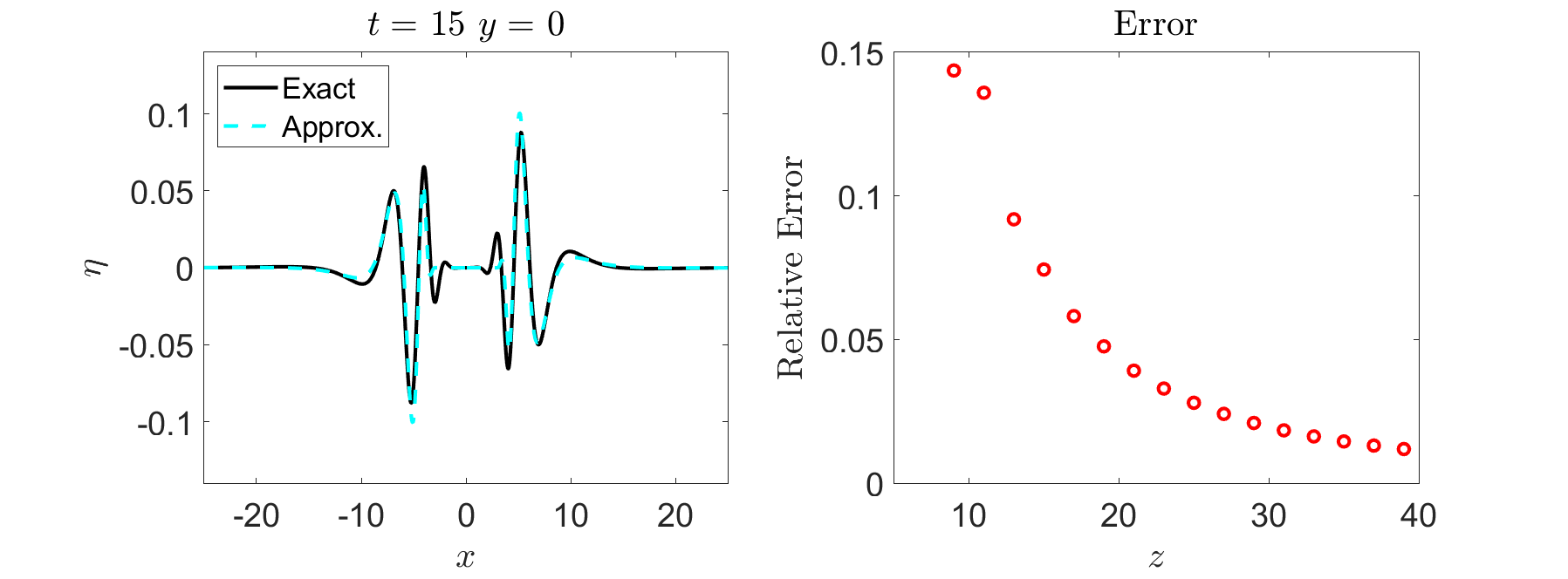}
    \caption{Comparison of the deep water (top left) exact solution \eqref{Eq: LinearGeneralOmega} computed 
    in terms of Fourier transforms 
    and the (top right) approximation given by the stationary phase formulae \eqref{Eq: InfiniteDepthApprox}. 
    (bottom left) Cross section at $y=0$ comparing the exact solution and approximation in the top row.
    (bottom right) Relative norm squared error \eqref{Eq: RelativeError} between the exact solution and the approximation is observed to converge like $O(t^{-2})$ as $t \rightarrow \infty$. Here $g=1$.}
        \label{Fig: InfiniteDepthApprox}
  \end{center}
\end{figure}

A comparison between the exact solution obtained from Eq.~\eqref{Eq: LinearGeneralOmega} with the deep water dispersion 
and the stationary phase approximation in Eq.~\eqref{Eq: InfiniteDepthApprox} is shown in Figure \ref{Fig: InfiniteDepthApprox}.
We observe that both qualitatively and quantitatively, the approximation describes the solution. To quantify how accurate the stationary phase approximation is, we compute the relative norm squared error
%
\begin{equation}
  \mathrm{Relative ~Error} = \frac{\iint \left|\eta_{\mathrm{Exact}} - \eta_{\mathrm{Approx}} \right|^2 \dd A}
            {\iint \left|\psi_{\mathrm{Exact}}  \right|^2 \dd A},
  \label{Eq: RelativeError}
\end{equation}
and observe that this converges like $O(t^{-2})$ as $t$ increases.
This second-order convergence rate resembles those observed in the context of Klein-Gordon spiral wave solutions found in \cite{AblCoNix2025}.

\subsubsection{Deep Water Spiral Structure}
\label{spiral_shape_sec}

\begin{figure}
  \begin{center}
    \includegraphics[width = 0.8\textwidth]{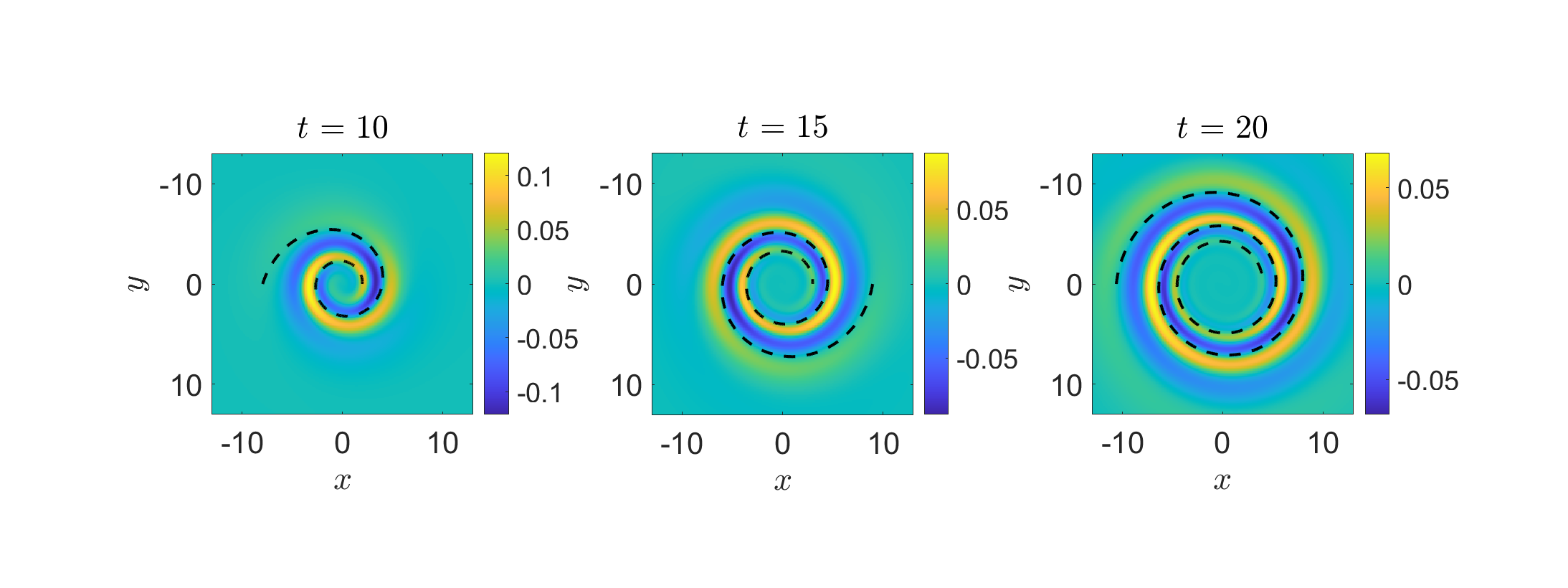}
    \caption{Comparison of the equation for the spiral shape \eqref{Eq: SpiralShape} found in the deep water limit (dotted line)
     and the exact solution computed by Fourier transform. The black dashed curve tracks the zeros of the spiral.}
         \label{Fig: SpiralShape}
  \end{center}
\end{figure}

The fundamental shape of the deep water spirals found above 
can be derived from Eq.~\eqref{Eq: InfiniteDepthApprox}. 
Since the surface spiral consists of regions of elevation (positive) and depression (negative), 
we can trace the shape of the spiral along the zeros of the solution, which occur at
\begin{equation}
  \frac{2 \ol{r}}{g} \sin(\theta) \cos\left(\frac{g t}{4 \ol{r}} \right) - \cos(\theta) \sin\left(\frac{g t}{4 \ol{r}} \right) = 0 
\end{equation}
or, in terms of the unscaled variables, 
\begin{equation}
  r \tan(\theta) = \frac{gt}{2} \tan\left(\frac{g t^2}{4 r} \right).
\end{equation}
%
We are left with an equation that cannot be solved explicitly. 
However, if we look for the zeros ($\theta = 0$) that lie along the positive the $x$-axis, 
we find that $n^{\mathrm{th}}$ zero has a radius of
\begin{equation}
  \frac{g t^2}{4 r_n} = n \pi, ~~~~~~~ n \in \mathbb{Z} .
\end{equation}
Interpolating these zeros on the x-axis for general $\theta$, it suggests that 
the shape of the spiral is given by 
\begin{equation}
  r = \frac{gt^2}{4 \theta}.
    \label{Eq: SpiralShape}
\end{equation} 
For a fixed time $t$, this is a hyperbolic spiral. 
A comparison of curve given in \eqref{Eq: SpiralShape} superimposed on top of the exact solution in Eq.~\eqref{Eq: LinearGeneralOmega}  
is shown in Figure~\ref{Fig: SpiralShape}. The curve fits the zeros of the spiral with good agreement between the two.
While it is not shown here, these curves correspond to the extrema 
of the surface velocity profile, $\eta_t(x,y,t)$. 
It is interesting to note that hyperbolic spirals have been used to describe certain galaxies 
where the radial spiral arm grows as it moves away from the center \cite{Kenn1981}. 

\section{Spiral waves: weakly nonlinear}
\label{Sec: weakNL}

Returning to the AFM formulation 
we consider the weakly nonlinear limit in more detail; 
letting $\eta \rightarrow \eps \eta$ and $q \rightarrow \eps q$, in \eqref{Eqs: EtaQEquatons} yields
\begin{subequations}
\begin{equation}
  \iint_{\mathbb{R}^2} d \bv{r} e^{ - i {\bf k} \cdot {\bf r} }
    \left( i \eta_t \cosh\Big[k h +\eps k \eta \Big] 
    -  \frac{{\bf k} \cdot \nabla q}{k}  \sinh \Big[k h +\eps k \eta \Big]\right)  = 0 \;,
      \label{Eq: EtaQEpsilon1}
\end{equation}
\begin{equation}
  q_t  + g\eta + \frac{\eps}{2} | \nabla q|^2 - \eps \frac{(\eta_t + \eps \nabla q \cdot \nabla \eta)^2} {2(1+ \eps^2 |\nabla \eta|^2)}=0\;.
    \label{Eq: EtaQEpsilon2}
\end{equation}
    \label{Eqs: EtaQEpsilons}
\end{subequations}
Expanding \eqref{Eq: EtaQEpsilon1} in powers of $\eps$, we find
\begin{align}
    0 = &~\ft{\eta}_t + \tanh\big(k h \big) \frac{\ii \bv{k}\cdot\ft{\grad q}}{k}  + \eps \Big(k \tanh\big(k h \big)  \ft{\eta \eta_t} + \ii \bv{k}\cdot \ft{\eta \nabla q} \Big) 
        \label{Eq: ImplicitEtaT} \\
    &\qquad \qquad   + \frac{\eps^2}{2} \Bigg(k^2  \ft{\eta^2 \eta_t}  +  k \Big( \ii \bv{k}\cdot \ft{\eta^2 \nabla q }\Big) \tanh\big(k h \big) \Bigg) + \ldots.\notag
\end{align}
In order to evolve the system, we need to solve for $\eta_t$. However, in this form, $\eta_t$ is nested within the nonlinear terms.
Thus, we must find $\eta_t$ iteratively,
\begin{equation}
    \eta_t = H_0 + \eps H_1 + \ldots. 
    \label{etat_expand}
\end{equation}
Substituting this into equation \eqref{Eq: ImplicitEtaT} and collecting powers of $\eps$ gives 
\SubAlign{
    H_0 & =  - \frac{1}{2\pi} \iint_{\mathbb{R}^2} d \bv{k} e^{  i {\bf k} \cdot {\bf r} } ~\tanh\big(k h \big) \frac{\ii \bv{k}\cdot\ft{\grad q}}{k} \, \\
    H_1 & =  - \frac{1}{2\pi}  \iint_{\mathbb{R}^2} d \bv{k} e^{  i {\bf k} \cdot {\bf r} } ~ \left[ k \tanh\big(k h \big)  \ft{\eta H_0} + \ii \bv{k}\cdot \ft{\eta \nabla q } \right] .
}[Eqs: EtaTExpansion]
Equation \eqref{Eq: ImplicitEtaT} may now be written in explicit terms as 
\begin{align}
    \ft{\eta}_t \approx &~  - \tanh\big(k h \big) \frac{\ii \bv{k}\cdot\ft{\grad q}}{k}  - \eps\Big( k \tanh\big(k h \big)  \ft{\eta H_0} + \ii \bv{k}\cdot \ft{\eta \nabla q }\Big) 
        \label{Eq: EtaSecondOrder}\\
    &\qquad  - \eps^2 \Bigg(k \tanh\big(k h \big)  \ft{\eta H_1} +  \frac{k^2}{2}  \ft{\eta^2 H_0}  + \frac{k}{2} \Big( \ii \bv{k}\cdot \ft{\eta^2 \nabla q }\Big) \tanh\big(k h \big) \Bigg)  \notag 
\end{align}
where we truncate the expansion at O($\eps^2$). 

Applying expansion \eqref{etat_expand} in \eqref{Eq: EtaQEpsilon2} now gives 
\SubAlign{
    0 & = q_t +  g\,\eta +  \frac{\eps}{2} \big|\grad q\big|^2  -  \frac{\eps}{2} \frac{\big(\eta_t + \eps \grad \eta \cdot \grad q\big)^2}{1+\eps^2\big| \grad \eta \big|^2} \\
    0 & = q_t +  g\,\eta + \frac{\eps}{2} \big|\grad q\big|^2 \notag\\
    &\qquad -  \frac{\eps}{2} \Big( \eta_t^2 + 2\eps \eta_t \grad \eta \cdot \grad q +  \ldots \Big)\Big(1 - \eps^2 \big| \grad \eta \big|^2 + \ldots \Big) \\
    0 & = q_t +  g\,\eta +  \frac{\eps}{2} \big|\grad q\big|^2  -  \frac{\eps}{2} \eta_t^2 - \eps^2 \eta_t \grad \eta \cdot \grad q +  \ldots \\
    0 & = q_t + g\,\eta +  \frac{\eps}{2} \big|\grad q\big|^2  -  \frac{\eps}{2} (H_0^2 + 2 \eps  H_0 H_1) - \eps^2 H_0\grad \eta \cdot \grad q - \cdots
}
where $H_0$ and $H_1$ are taken from the $\eta_t$ expansion \eqref{Eqs: EtaTExpansion}. Hence,
\begin{equation}
    q_t \approx  - g\,\eta + \frac{\eps}{2} \Big(H_0^2 - \big|\grad q\big|^2 \Big) + \eps^2 \Big( H_0 H_1 +  H_0 \grad \eta \cdot \grad q \Big) .
\end{equation}

The final system describing the evolution of  weakly nonlinear waves is given by
\SubAlign{
     \ft{\eta}_t &\approx - \tanh\big(k h \big) \frac{\ii \bv{k}\cdot\ft{\grad q}}{k}  - \eps\Big( k \tanh\big(k h \big)  \ft{\eta H_0} + \ii \bv{k}\cdot \ft{\eta \nabla q }\Big) \notag \\
    &\qquad  - \eps^2 \Bigg(k \tanh\big(k h \big)  \ft{\eta H_1} + \frac{k^2}{2}  \ft{\eta^2 H_0}  +  \frac{k}{2} \Big( \ii \bv{k}\cdot \ft{\eta^2 \nabla q }\Big) \tanh\big(k h \big) \Bigg)\\
    \left(\ft{\grad q}\right)_t &\approx  \ii \bv{k}   \Bigg(- g\,\ft{\eta} + \frac{\eps}{2} \Big(\ft{H_0^2} -\ft{\big|\grad q\big|^2} \Big) + \eps^2 \Big( \ft{H_0 H_1} +  \ft{H_0\grad \eta \cdot \grad q} \Big) \Bigg).
}[Eq: NonlinearWaterWaves]
where $H_0,H_1$ are given by \eqref{Eqs: EtaTExpansion}. Or, in terms of the dispersion relation, $\omega(k)$, we can write these equations as 
\SubAlign{
      H_0 & =  - \frac{1}{2\pi g} \iint_{\mathbb{R}^2} d \bv{k} e^{  i {\bf k} \cdot {\bf r} } ~\omega^2(k) \frac{\ii \bv{k}\cdot\ft{\grad q}}{k^2} \, \\
    H_1 & =  - \frac{1}{2\pi g}  \iint_{\mathbb{R}^2} d \bv{k} e^{  i {\bf k} \cdot {\bf r} } ~ \left[\omega^2(k) \ft{\eta H_0} + \ii g \bv{k}\cdot \ft{\eta \nabla q } \right]\\
    \ft{\eta}_t &\approx - \omega^2(k) \frac{\ii \bv{k}\cdot\ft{\grad q}}{g k^2}  - \eps\Bigg( \frac{1}{g}\omega^2(k)   \ft{\eta H_0} + \ii \bv{k}\cdot \ft{\eta \nabla q }\Bigg) \notag \\
    &\qquad  - \eps^2 \Bigg(\frac{1}{g}\omega^2(k) \ft{\eta H_1} + \frac{k^2}{2}  \ft{\eta^2 H_0}  +  \frac{1}{2g}\omega^2(k) \Big( \ii \bv{k}\cdot \ft{\eta^2 \nabla q }\Big) \Bigg)\\
    \left(\ft{\grad q}\right)_t &\approx  \ii \bv{k}   \Bigg(- g\,\ft{\eta} + \frac{\eps}{2} \Big(\ft{H_0^2} -\ft{\big|\grad q\big|^2} \Big) + \eps^2 \Big( \ft{H_0 H_1} +  \ft{H_0\grad \eta \cdot \grad q} \Big) \Bigg).
}[Eq: NonlinearWaterWavesAlt]

In these equations we evolve $\ft{\eta}$, $\ft{q_x}$ and $\ft{q_y}$ (the final line is vectorial and contains two equations in it). 
We point out that  the last line has $\nabla q$. 
This has two benefits, one physical and one numerical. First, $\nabla q$ represents the surface fluid velocity and unlike $q$ is physical.
Second, the dominant source of numerical instability arises in the high frequency ($k\gg 1$) modes, and in this form the scaling of the nonlinear terms 
with respect to $k$ is reduced in order, i.e., terms of order $\eps$ scale like O($k$) instead of O($k^2$) 
and terms of order $\eps^2$ scale like O($k^2$) instead of O($k^3$).

The numerical scheme used to solve \eqref{Eq: NonlinearWaterWaves} is a 4th-order Runge-Kutta method with integrating factor. 
Details are given in the Appendix.

\begin{figure}
  \begin{center}
    \includegraphics[width = 0.8\textwidth]{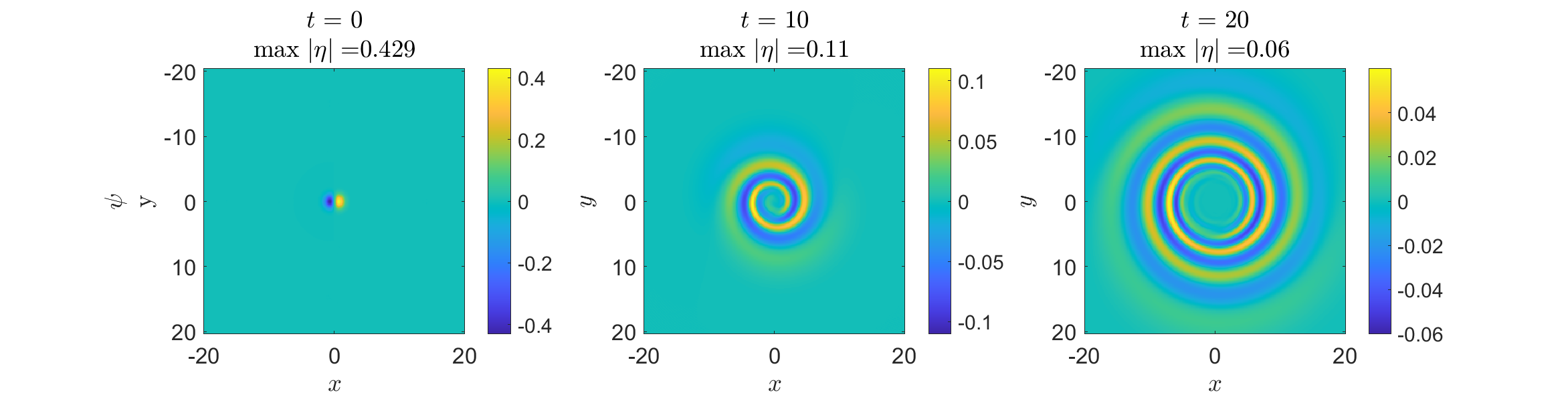}
    \caption{Evolution of the nonlinear water wave equations \eqref{Eq: NonlinearWaterWaves} with initial conditions \eqref{Eq: SpiralIC}. 
    Here $\omega$ is given for finite depth waves by \eqref{Eq: OmegaFiniteDepth}, $\eps = 0.25$, $h=0.7$ and $g=1$. 
    Compare this with the linear results shown in Fig.~\ref{Fig: FiniteDepthAsymp}.}
        \label{Fig: NLEvolutionFinitDepth}
  \end{center}
\end{figure}

Using the same initial conditions \eqref{Eq: SpiralIC} prescribed in the linear system described Section \ref{linear_spiral_sec}, 
we find that linear spiral waves constitute a robust family solutions. 
Adding weak nonlinearity does not have significant impact on the formation and persistence of the spiral waves observed in the linear problem.
Snapshots of the nonlinear evolution for finite depth waves are shown in Fig.~\ref{Fig: NLEvolutionFinitDepth}. 
These results can be compared against the linear evolution shown in Fig.~\ref{Fig: LinearEvolution}. The difference between the two is minor
with the nonlinear version showing a few extra spirals near the center and a slightly larger peak amplitude.

\begin{figure}
  \begin{center}
    \includegraphics[width = 0.8\textwidth]{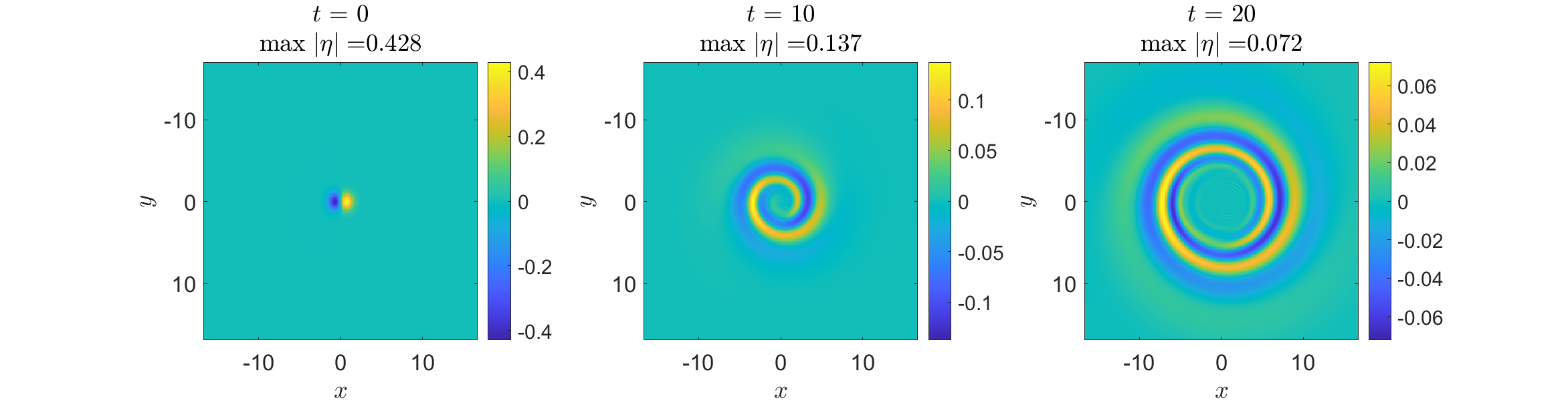}
    \caption{Evolution of the nonlinear water wave equations \eqref{Eq: NonlinearWaterWaves} with initial conditions \eqref{Eq: SpiralIC}. 
    Here $\omega$ is given for infinite depth waves by \eqref{Eq: OmegaDeepWater}, $\eps = 0.25$, and $g=1$. Compare this with the linear results shown in Fig.~\ref{Fig: InfiniteDepthApprox}.}
        \label{Fig: NLEvolutionInfinitDepth}
  \end{center}
\end{figure}

Next, the deep water spirals in the presence of weak nonlinearity is highlighted in Fig.~\ref{Fig: NLEvolutionInfinitDepth}, 
taking the dispersion relation $\omega^2(k) = g k$. These results can be compared with the linearized version in Fig.~\ref{Fig: InfiniteDepthApprox}. 
Comparing the two at $t = 10$, we see that there is little difference in the shape, so the hyperbolic spiral in discussed in Sec.~\ref{spiral_shape_sec} is 
a good description. Similar to finite depth, the nonlinear solution has a slightly larger amplitude than the linear version. 
For the cases considered here, the weak nonlinearity has modest impact on the evolution of the linear spiral waves. 
We also find numerically that adding a small amount of noise to the initial conditions has little effect; i.e. the spiral waves appear to be stable.

\section{Conclusion}

Spiral waves are found in the linear and weakly nonlinear irrotational water wave equations. 
These unsteady waves evolve from a suitable class of initial conditions; there is no external forcing. 
In the linear problem, long-time asymptotic approximations via stationary phase are found to be in good 
agreement with the exact solution. 
Numerical simulations indicate that unsteady spiral solutions are maintained for weak nonlinearity. 
While spiral waves are well-known in excitable media modeled by reaction-diffusion equations, they apparently have not been previously 
analyzed in fundamental linear or nonlinear 
irrotational water waves. As such, these solutions are new  two space one time dimensional
solutions to the classical water waves equations.


\section{Acknowledgments}
MJA was partially supported by NSF under Grant No. DMS-2306290.

\section{Declaration of Interests}
The authors report no conflict of interest

\bibliographystyle{unsrt}
\bibliography{WaterWaveSpiralBib}

\appendix
\section{Numerical Simulations}

The numerical evolution of the nonlinear water wave equations \eqref{Eq: NonlinearWaterWaves} is 
stiff due to the multiplication by nonlinear Fourier wavenumbers at high frequency ($k\gg 1$).
This restriction can be reduced by introducing an integrating factor that moves 
the linear component into a bounded exponential factor \cite{Tre00}. Observe that the Fourier transform of the linear part of \eqref{Eq: NonlinearWaterWaves} can be 
written as
\begin{equation}
  \pdv{}{t}
    \begin{bmatrix}
      \ft{\eta} \\ \ft{q_x} \\ \ft{q_y} 
    \end{bmatrix}
  = L 
    \begin{bmatrix}
      \ft{\eta} \\ \ft{q_x} \\ \ft{q_y} 
    \end{bmatrix}
  ,\qquad \mathrm{where}\quad 
  L = 
    \begin{bmatrix}
      0& -\frac{\ii k_x \omega^2(k) }{g k^2} & -\frac{\ii k_y \omega^2(k) }{g k^2}\\
      -\ii g k_x & 0 & 0 \\ 
      -\ii g k_y & 0 & 0
    \end{bmatrix}.
\end{equation}
The eigenvalues of $L$ are $0 , \pm i \omega(k)$ and have no nonzero real-part. 
As a result, arbitrary solutions of the linear system remain bounded for all time.
The integrating factor is given by
\begin{equation}
 \ee^{L t} =
      \begin{bmatrix}
      \cos\big(\omega t \big)& -\frac{\ii k_x \omega \sin\big(\omega t \big)}{g k^2 } & -\frac{\ii k_y \omega \sin\big(\omega t \big)}{g k^2 }\\
      -\frac{\ii g k_x\sin\big(\omega t \big)}{\omega} & \frac{k_y^2 + k_x^2  \cos\big(\omega t \big)}{k^2} & - \frac{k_x k_y \left[1-\cos\big(\omega t \big) \right]}{k^2} \\ 
      -\frac{\ii g k_y\sin\big(\omega t \big)}{\omega} & - \frac{k_x k_y \left[1-\cos\big(\omega t \big)\right]}{k^2}  & \frac{k_x^2 + k_y^2  \cos\big(\omega t \big)}{k^2} 
      \end{bmatrix} . 
      \notag
\end{equation}
The evolution of the of the nonlinear wave equation \eqref{Eq: NonlinearWaterWaves} 
employs a 4th-order Runge Kutta method in $\bv{v}$ where
\SubAlign{
    H_0 &\equiv - \frac{1}{2\pi} \iint_{\mathbb{R}^2} \ee^{\ii \bv{k}\cdot \bv{r}}  ~   \tanh\big(k h \big) \frac{\ii \bv{k}\cdot\ft{\grad q}}{k} \, \dd \bv{k} \\
    H_1 &\equiv - \frac{1}{2\pi} \iint_{\mathbb{R}^2} \ee^{\ii \bv{k}\cdot \bv{r}}  ~  \Bigg(k \tanh\big(k h \big) \ft{\eta H_0} + \ii \bv{k}\cdot \ft{\eta \nabla q }  \Bigg)\, \dd \bv{k}\\
      \begin{bmatrix}
        \ft{\eta} \\ \ft{q_x} \\ \ft{q_y} 
      \end{bmatrix} 
        & =  \ee^{L t}\bv{v}(t) \\
    M_1 & =  - \eps\Big( k \tanh\big(k h \big)  \ft{\eta H_0} + \ii \bv{k}\cdot \ft{\eta \nabla q }\Big) \notag \\
    &\qquad  - \eps^2 \Bigg(k \tanh\big(k h \big)  \ft{\eta H_1} + \frac{k^2}{2}  \ft{\eta^2 H_0}  +  \frac{k}{2} \Big( \ii \bv{k}\cdot \ft{\eta^2 \nabla q }\Big) \tanh\big(k h \big) \Bigg)\\
    M_2 & =  \ii k_x \Bigg(- g\,\ft{\eta} + \frac{\eps}{2} \Big(\ft{H_0^2} -\ft{\big|\grad q\big|^2} \Big) + \eps^2 \Big( \ft{H_0 H_1} +  \ft{H_0\grad \eta \cdot \grad q} \Big) \Bigg)\\
    M_3 & =  \ii k_y \Bigg(- g\,\ft{\eta} + \frac{\eps}{2} \Big(\ft{H_0^2} -\ft{\big|\grad q\big|^2} \Big) + \eps^2 \Big( \ft{H_0 H_1} +  \ft{H_0 \grad \eta \cdot \grad q} \Big) \Bigg)\\
    \bv{v}_t &= \ee^{-L t}
      \begin{bmatrix}
        M_1 \\ M_2 \\ M_2 
      \end{bmatrix}.
}

We find that this integrating factor approach allows us to take larger time steps by an order of magnitude, significantly speeding up these 2+1 dimensional simulations.

\end{document}